\documentclass{ws-tpe}
\usepackage{multicol}
\usepackage{float}
\usepackage{booktabs}
\usepackage{graphicx}
\begin{document}

\markboth{Lachlan McGinness}{The benefits and challenges of a Quantum Computing Concept Inventory}

\title{The Benefits and Challenges of a Quantum Computing Concept Inventory}

\author{Lachlan M\textsuperscript{c}Ginness}

\address{School of Computing, Australian National University\\
Acton, ACT, Australia\\
\email{lachlan.mcginness@anu.edu.au}}

\maketitle


\begin{abstract}
A Quantum Computing Concept Inventory is needed for the acceleration of uptake of best practice in quantum computing education required to support the quantum computing workforce for the next two decades. Eight experts in quantum computing, quantum communication or quantum sensing were interviewed to determine if there is substantial non-mathematical content to warrant such an inventory and determine a preliminary list of key concepts that should be included in such an inventory. Developing such an inventory is a challenging task requiring significant international `buy-in' and creativity to produce jargon-free valid questions which are accessible to students who are yet to study quantum mechanics.
\end{abstract}


\begin{multicols}{2}
\section{Introduction}

Quantum Computing, quantum sensing and quantum communication are rapidly advancing fields which will have a significant economic and industrial impact globally\cite{Foley2020}. Australia's Chief Scientist and the Commonwealth Science and Industry Research Organisation (CSIRO) released a report stating the anticipated impact of quantum technologies on Australia's industries\cite{Foley2020}. The report predicts over \$4 billion of revenue and the creation of over 16,000 jobs by 2040\cite{Foley2020}\textsuperscript{,}\cite{CSIRO2022}. 

Where will these workers for the quantum industry come from? Currently the majority of the people who work in quantum research and development obtained PhDs in the field\cite{SQA2023}. However in Australia there are only 10,000 PhD students per year across all disciplines and the vast majority of these do not study quantum computing\cite{McCarthy2023}\textsuperscript{,}\cite{VU2024}. Therefore it is not feasible to produce 16,000 quantum computing PhD's by 2040. In order to meet the workforce demands people will need to be trained through alternate pathways. 

When governments and companies aim to hire new employees how will they determine that they have the required skills? One possible system is a qualification-based system where anyone who has obtained, say, a masters degree or diploma is assumed to have the required skills to work in the industry. 

Another approach would be to have a standardised test that could be used to measure understanding of quantum phenomena. These standardised tests could aim to measure a combination of mathematical and conceptual understanding similar to a traditional physics exam. However if one is interested in measuring only conceptual understanding of quantum computing then it may be more appropriate to use a Concept Inventory.

\subsection{Concept Inventories}
In 1992 Hestenes, Wells and Swackhammer revolutionised Physics Education Research (PER) with the invention of the Force Concept Inventory (FCI)\cite{FCI}. The FCI was a multiple choice test that was purely conceptual; mathematics was not required to answer the questions. The results surprised many physics instructors as they revealed that students do not understand many basic concepts about Newton's laws that were assumed to be already mastered. The FCI and a similar instrument called the Force and Motion Concept Evaluation (FMCE) quickly became gold standards in measuring student understanding\cite{FMCE}. By giving students the test both at the start and end of a physics course, physics educators could measure the effectiveness of their teaching and the success of any education initiatives that they implemented. 

The Force Concept Inventory's success can be attributed to a few factors. Firstly, there was a well agreed upon set of concepts (Netwon's Laws) that many different instructors were interested in teaching to their students across institutions. Secondly, the FCI questions are about everyday life scenarios and do not use jargon and therefore are accessible to students who have not studied the topic before. Finally, the instrument was used across many institutions allowing for the comparison of different teaching initiatives in different settings. 

The success of the Force Concept Inventory led to an explosion of Concept Inventories and Conceptual Tests being developed for different concepts in physics\cite{McGinness2016}\textsuperscript{,}\cite{Aslanides2013} and even just in the niche field of quantum mechanics\cite{QMCA}\textsuperscript{,}\cite{QMS}\textsuperscript{,}\cite{QPCS}\textsuperscript{,}\cite{QMVI}\textsuperscript{,}\cite{QMFPS}\textsuperscript{,}\cite{Madsen2017}. In order to ensure that the developed instruments were of high quality, Wendy Adams and Nobel Laureatte Carl Wieman published a procedure for the effective creation of conceptual inventories outlining the following phases\cite{Adams2011}:
\begin{enumerate}
	\item Consulting international experts and leaders in the field to determine the list of important concepts and definitions.
	\item Determining a draft list of questions.
	\item Performing think-aloud interviews with students to determine the validity of the questions and use students' incorrect answers as a source of distractors.
	\item Giving the test to a large number of students in order to check the reliability and validity of the test overall as well as the individual questions.
\end{enumerate}

Although many quantum-based concept inventories have been developed since the publication of this process, none of them have had an impact even close to that of the FCI. We can attribute a few causes to this:
\begin{itemize}
	\item The FCI is already an existing standard that allows teaching initiatives to be tested on large classes (generally mechanics is one of the topics covered in first year classes). There is less incentive to test teaching initiatives on niche subjects were class sizes are smaller and statistical significance is less likely to be achieved. 
	\item Many of the inventories that have been developed have not rigorously followed the process outlined by Adams and Wieman. For example, rather than following the long and difficult process of consulting international experts, a faculty meeting was held to decide on the concepts for some inventories\cite{QMCA}. These shortcuts result in the inventory targeting a very specific set of topics that are covered in the author's course rather than a list of concepts which are agreed to be important by the PER community.
	\item Many inventories use jargon that a student would not be able to understand without first having studied the course. This drastically reduces the validity of pretests and prevents the inventory from being used as a valid measurement of the effectiveness of teaching. It also makes the experience of doing the pretest very negative for students.
\end{itemize}

\subsection{Quantum Computing Concept Inventory}
There are significant challenges in implementing a Quantum Computing Concept Inventory. Firstly, one would need to be very creative in developing questions that were able to test underlying concepts without using jargon. Secondly for the inventory to provide a useful benchmark, significant uptake would be required from the PER and quantum computing education communities. Therefore, one would need to get buy-in from a range of international experts and programs that are aiming to teach quantum computing.

Despite these challenges, there is still a strong case for the development of a Quantum Computing Concept Inventory. Firstly the quantum workers of 2040 are currently in primary school or high school and there are a number of programs such as the Sydney Quantum Academy\cite{SQA2024}, Einstein First\cite{Einsteinian2024}, and the Coding School's Qubit by Qubit program\cite{QxQ} which are attempting to teach core quantum computing concepts to these students without high levels of mathematics. These programs are understandably keen to assess the effectiveness of their teaching and improve their programs. If a Concept Inventory is not available, then institutions like these would each be required to develop their own instruments. Given the significant effort of creating a rigorous conceptual test, it is likely that these tests will contain jargon and measure a set of concepts that are specific to the course instruction. The lack of a common benchmark will make it impossible for different institutions to compare their evaluations.

Rigorous PER would accelerate the uptake of best teaching practice in this new teaching area. This new area of PER would need to be united and guided by a tool like a Quantum Computing Concept Inventory with:
\begin{itemize}
	\item concepts determined by international leaders in the field from across institutions and countries.
	\item buy-in from the entire quantum computing education community.
	\item non-mathematical, jargon-free, conceptual questions that are accessible to students who have yet to start their quantum studies.
\end{itemize}

Achieving each of these criteria requires significant engagement with a large number of international leaders in the field. In this paper we present the findings from a preliminary study which has contacted only a small number ($n=8$) of international experts. We present a breakdown of the demographics of the consulted experts, the concepts that they think would be required for inclusion and analyse their first ideas for questions to be included. 

\section{Survey Respondents}
Eight quantum computing, quantum sensing and quantum communication experts were interviewed for this study. To be classified as an expert the person needed to be currently teaching a quantum computing course or be a professional researcher of quantum computing, quantum communication or quantum sensing. Four experts were from Australia, two from Singapore and two from the United States of America. Three of the respondents worked in the university sector, four for government funded agencies and one for private enterprise. Six of the interviewees were male and two were female. Table 1 
explains the quantum background of the experts.

Interviews were carried out in accordance with the ethics approval from the Australian National University Human Research Ethics Council (Protocol 2023/097). In addition to the demographic information above, experts were asked the following questions:
\begin{enumerate}
	\item If you had to break your knowledge of quantum computing/sensing into mathematical and conceptual, what percentage would be in each?
	\item What are the key conceptual ideas that you use in your work/that you teach students in your course?
	\item Do you have any suggestions for questions that you could use to test students' understanding of these concepts?
\end{enumerate}

\vspace{-0.5cm}

\begin{tablehere}
	\tbl{Work experience areas of the eight interviewed experts. Note most experts stated that they have experience in multiple of these areas.}
	{\begin{tabular}{@{}cc@{}}
			\toprule
			Experience Area & Number of Experts\\ \colrule
			Quantum Research & 8 \\
			Quantum Industry & 3 \\
			Taught a Quantum Course & 4 \\
			Worked in Theory & 4 \\
			Worked in Experiment & 5 \\
			Worked in Software & 1 \\
			Hardware (Qubits and Gates) & 5 \\
			Hardware (Sensing + other) & 3 \\
			Protocols and Algorithms & 3 \\
			\botrule
	\end{tabular}}\label{tab:quanum_experience}
\end{tablehere}

\vspace{-0.5cm}

The answers to these questions are summarised in the following section.

\section{Results from Interview Questions and Discussion}

The first question was asked in order to help the experts make a distinction between mathematical knowledge and conceptual understanding. Respondents found the exercise of trying to split their knowledge both interesting and challenging. One respondent said that the task was impossible and refused. The results from the remaining seven respondents are displayed in Figure \ref{fig:hist}.

\begin{figurehere}
	\centerline{
		\includegraphics[width=0.5\textwidth]{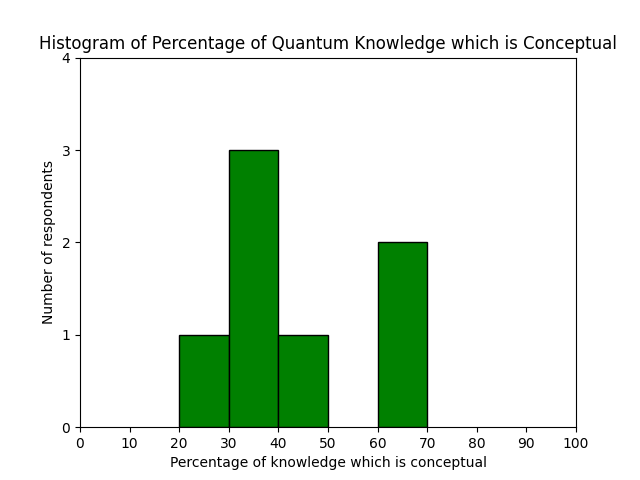}
	}
	\caption{Histogram of amount of quantum knowledge which is conceptual rather than mathematical. The one expert who said that they were unable to disentangle their mathematical and conceptual quantum knowledge is not included.}
	\label{fig:hist}
\end{figurehere}

The results from Figure \ref{fig:hist} show that there is a significant amount of conceptual content which could be taught even if students don't have the mathematical requirements to perform quantum mechanical calculations. This provides validity to the proposal that a Quantum Computing Concept Inventory is both possible and useful. 

There is significant variation between the perceived amount of mathematical and conceptual knowledge between experts. We note that these numbers are just estimates which could lead to some variation. It is also possible that in different roles different amounts of mathematical knowledge is used. The responses between 21\%-30\% and 61\%-70\% were given by experimentalists. 

\subsection{Key Quantum Concepts}

In response to the second question, most experts had many different concepts that they considered core to understanding quantum computing, quantum sensing or quantum communication. Concepts that were mentioned by at least two experts are shown in Table 2.

Concepts that were only mentioned once include: Adiabatic, Annealing, Classical System, Computation, Deterministic, Energy, Entropy, Excitation, Exotic particles, Noise, Pure State, Quantum Algorithm, Quantum Contextuality, Quantum Circuit, Quantum information, Quatum Non-commutative Operators, Quantun Non-locality, Quantum Simulation, Quantum Supremacy, Quantum Teleportation, Spin, Tunnelling, Uncertainty Principle, Universal Quantum Computer and Wavefunction. Although all of these concepts are relevant, not all concepts can be considered core, only those that the majority agree on.

\vspace{-0.3cm}

\begin{tablehere}
	\tbl{List of concepts mentioned by at least two respondents during interviews.}
	{\begin{tabular}{@{}cc@{}}
			\toprule
Concept & Number of Experts\\ \colrule
Coherence & 2 \\
Decoherence &2 \\
Entanglement & 5 \\
Interference & 2 \\
Measurement & 3 \\
No Cloning Theorem & 2 \\
Probability & 2 \\
Quantum State &2 \\
Superposition & 5 \\\botrule
	\end{tabular}} 
\end{tablehere}

\vspace{-0.2cm}

Table 2 
contains 10 concepts which is a good number for an inventory. Note Coherence and Decoherence should probably be combined to form a single concept. 

We propose that before finalising any list, a large number of experts (ideally $n \approx 50$) from a range of institutions and countries should be surveyed. This will ensure that the concepts are considered core by a larger field of experts. However, this list could serve as a starting point for such a survey. 

It is important to have a strong definition of each concept when asking for expert agreement . When experts were asked to give definitions for these concepts, answers varied significantly. The following two examples were both definitions given for entanglement, however one is strongly mathematically focused while the other is more conceptually focused:
\begin{itemize}
	\item \textit{``An entangled state is a pure state that cannot be factorised.''}
	\item \textit{``An entangled state is where two or more components of a quantum system are dependent and measuring one part of the system gives you accurate information about the other.''}
\end{itemize}

Although mathematical definitions are important, for a Concept Inventory we would like to measure the students' conceptual understanding and therefore we would recommend conceptually-based definitions more similar to the second. 

\subsection{Potential Questions}

The questions recommended by the experts fall into a number of categories. Many of the questions recommended were either `Explain this concept...' or `List the differences between...'. Both of these types of questions could be used in think-aloud interviews with students in order to determine a list of multiple choice options to be used as Quantum Computing Concept Inventory questions.

However, we recommend against such questions. Better Concept Inventory questions are questions that relate to a physical outcome. This way the choice of keywords or model for the phenomenon is not judged. If the student chooses a wrong answer it should be because they are predicting something that is provably incorrect. 

Another example of a question suggested by an expert is:\\ 

\textit{Is the following state entangled?} $ | 0, 1 \rangle + | 1, 0 \rangle $\\

This could be easily turned into a multiple choice question by giving two options and then stating which is entangled, A, B, both or neither. However we would argue that such a question is not a good choice for a Concept Inventory as it relies on the reader having seen bra-ket notation before.  Instead we recommend questions which do not rely on mathematical notations or jargon.

Given these constraints, it will take significant creativity in order to develop Concept Inventory questions. We found that it takes experts a significant amount of time to develop questions that meet these requirements. Therefore, it is unlikely that they will develop good questions in the duration of an interview.

If there are any physical systems that are analogous to quantum systems, these could be a good starting point for Concept Inventory questions. An example of a starting point for a superposition/measurement question could be as follows:

\textit{Suppose an electron is stuck in a box and can exist in three states: A, B or C. Before a measurement is made what is true about the electron's state?
\begin{enumerate}
	\item[a)] The electron could be in state A, B or C, we just don't know which until the point of measurement. Once it is measured we will know which state it was in all along.
	\item[b)] The electron could be in a combination of states A, B and C at the same time. Once the measurement is made then the electron will be in one of the states.
	\item[c)] The electron could be constantly changing between states A, B and C with different probabilities. The time at which we measure the particle will determine which outcome we get. Once the measurement is made then the electron will be in one of the states.
\end{enumerate}}

One might read this question and think that it focuses on measurement rather than superposition, but note that the variation between the options does not correspond to the impact of the measurement but instead the state before measurement. In particular the distractor options are designed to appeal to the misconception that objects can only be in one state at one time. 

Both a and c are falsifiable by experiment, we can show that the outcome measured is independent of the time of measurement and double slit and similar experiments can show that particles are actually in multiple states at the same time (superposition is not just a lack of knowledge on our part). If the word `electron' is considered jargon then the word `particle' could be used instead. 

\section{Conclusion}

Given the urgency to develop a quantum-skilled workforce, acceleration of best teaching practice in quantum computing is essential. One way that this could be achieved is through the development of a Quantum Computing Concept Inventory which would allow instructors to measure the effectiveness of their instruction using a widely recognised benchmark. This would encourage publication of results which are comparable between institutions. 

Eight quantum experts were interviewed and the results show that there is substantial conceptual content that students can learn before studying the mathematics required to perform calculations. A preliminary list of important concepts includes: coherence/decoherence, entanglement, interference, measurement, the no cloning theorem, probability, quantum state and superposition.

Developing a Quantum Computing Concept Inventory will be challenging because it will require significant buy-in and contribution from the international quantum research and education communities, rigorous statistical testing in a large quantum computing course and significant creative thinking to find jargon-free, non-mathematical questions that are based in experimental outcomes rather than the student's choice of mental model.

\section*{Ethics}

The ethical aspects of this research have been approved by the ANU Human Research Ethics Committee (Protocol 2023/097).

\end{multicols}
\end{document}